\documentclass[a4paper,12pt]{article}

\date {}

\usepackage{amsfonts}
\usepackage{amsmath}

\newcommand{\be}{\begin{eqnarray}}
\newcommand{\ee}{\end{eqnarray}}

\def\Io{{\mathbb I}}

{}
{}
{}
{}
{}

\begin{document}
\title{Counting operator analysis of the discrete spectrum of some model Hamiltonians}
\author{Jan Naudts,
Tobias Verhulst\footnote{Research Assistant of the Research Foundation - Flanders (FWO - Vlaanderen)},
and Ben Anthonis\\
\strut\\
\small Departement Fysica, Universiteit Antwerpen,\\
\small Groenenborgerlaan 171, 2020 Antwerpen, Belgium\\
\small E-mail Jan.Naudts@ua.ac.be, Tobias.Verhulst@ua.ac.be
}
%\date {February 2009}
\maketitle

\begin{abstract}
The first step in the counting operator analysis of
the spectrum of any model Hamiltonian $H$ is the choice of
a Hermitean operator $M$ in such a way that the third commutator with $H$
is proportional to the first commutator. Next one calculates operators $R$ and $R^\dagger$
which share some of the properties of creation and annihilation operators,
and such that $M$ becomes a counting operator.
The spectrum of $H$ is then decomposed into multiplets,
not determined by the symmetries of $H$, but by those of
a reference Hamiltonian $H_{\rm ref}$, which is defined by
$H_{\rm ref}=H-R-R^\dagger$, and which commutes with $M$.
Finally, we introduce the notion of stable eigenstates. It is shown that
under rather weak conditions one stable eigenstate can be
used to construct another one.

\end{abstract}



\vskip 0.4cm

In the literature many attempts are found to generalise the notion of creation
and annihilation operators.
Some of these were introduced in the context of Bogoliubov's notion of quasi-particles
 --- see for instance \cite {KD65}. Others are related to the method developed by
Darboux in the nineteenth century to find new solutions of non-linear equations
(see for instance \cite {WC08}). Lowering and raising operators \cite {KB66} determine
recurrence relations and generate a Lie algebra (see for instance \cite {FLP01}).
The operators $R$ and $R^\dagger$ introduced below share some of the properties of
these operators.
In the case of the harmonic oscillator they coincide with the usual annihilation
and creation operators, up to some additive constants.
But in other models they can be used to guarantee the existence of a second eigenstate,
with a different eigenvalue. When applying the technique to a 4-site
Hubbard ring \cite {SR02,AVN08},
the spectrum of eigenvalues decomposes into multiplets, some of which have a fermionic appearance,
in contrast with the bosonic spectrum of the harmonic oscillator.

The one-dimensional Hubbard model has become very famous by the work of Lieb and Wu
\cite {LW68,LW03}
who used the Bethe ansatz to study the eigenvalues in the thermodynamic limit.
On the other hand, the energy levels of the 6-site Hubbard ring could be determined in a
reliable way \cite {HL71} by making use of the symmetries of the Hamiltonian.
In the latter work, energy level crossings were observed
which seemingly were not in agreement with the non-crossing rule of Wigner and von Neumann \cite {NW29}.
Their origin was explained in \cite {YAS02}. The multiplets considered here and
in \cite {AVN08}
are not determined by the symmetries of $H$ but by those of the reference
Hamiltonian $H_{\rm ref}=H-R-R^\dagger$.
In particular, we do not require integrability of the model Hamiltonian $H$.
For instance, our results are also applicable to the Hubbard model in dimensions larger than one.

Consider a quantum Hamiltonian $H$ together with a Hermitean operator $M\not=0$,
not commuting with $H$, such that
\be
[[[H,M],M],M]&=&\gamma^2 [H,M],
\label {model:ccr}
\ee
with real $\gamma\not=0$ \cite{foot1}.
Note that such an operator $M$ does always exist.
Indeed, let $M=E$ where $E$ is any
orthogonal projection operator (i.e.~$E=E^\dagger=E^2$).
Then (\ref {model:ccr}) is satisfied with $\gamma=1$,
as is readily verified.
The present work generalises this observation.

The relation (\ref {model:ccr}) is trivially satisfied if
\be
[[H,M],M]=\gamma^2 H.
\label {model:triv}
\ee
This is the case of the Pauli spin. The Hamiltonian is $H=\frac 12\hbar\omega_0\sigma_z$.
A short calculation gives $[[H,\sigma_x],\sigma_x]=4H$.
Hence, (\ref {model:ccr}) is satisfied with $M=\sigma_x$ and $\gamma=2$.

Less trivial are the examples where
\be
[[H,M],M]=\gamma^2 H+\mbox{ terms commuting with }M.
\label {model:lesstriv}
\ee
Note that (\ref {model:ccr}) implies (\ref {model:lesstriv}). Indeed,
one can always write $[[H,M],M]=\gamma^2 H+X$. Using (\ref {model:ccr}) there follows
\be
[X,M]&=&[[[H,M],M]-\gamma^2 H,M]\crcr
&=&[[[H,M],M],M]-\gamma^2 [H,M]\crcr
&=&0.
\ee

Consider the harmonic oscillator. The Hamiltonian is
\be
H=\frac 12\hbar\omega (bb^\dagger+b^\dagger b)
\ee
with $[b,b^\dagger]=1$. Let $M$ be the Hamiltonian of the shifted oscillator
\be
M=H-\zeta(b+b^\dagger)+\frac {\zeta^2}{\hbar\omega}.
\label {harmosc_countop}
\ee
Then one finds
\be
[[H,M],M]=(\hbar\omega)^2(H-M)-\zeta^2\hbar\omega.
\ee
This is of the form (\ref {model:lesstriv}) with $\gamma=\hbar\omega$.
Note that the choice of counting operator for this example differs from
the obvious one, this is, $M$ is not equal to $b^\dagger b$. The counting operator
$b^\dagger b$ commutes with the Hamiltonian $H$ and is therefore not
suited for the present analysis. In the one-parameter family (\ref {harmosc_countop}) of
allowed counting operators only the choice $\zeta=0$ is exceptional
because then $M$ commutes with $H$.

Also the Jaynes-Cummings model (see for instance \cite {RJC99,BHN02}) satisfies (\ref {model:lesstriv}).
The Hamiltonian is $H=H_0+V$ with
\be
H_0&=&\frac 12\hbar\omega (bb^\dagger+b^\dagger b)+\frac{1}{2}\hbar\omega_0\sigma_z\\
V&=&\hbar\kappa\big( b^\dagger\sigma_-+ b\sigma_+\big).
\ee
Take $M=H_0$. A short calculation gives
\be
[[H,M],M]=\hbar^2(\omega_0-\omega)^2(H-M).
\ee
Hence, (\ref {model:ccr}) is satisfied with $\gamma=\hbar(\omega_0-\omega)$
(assuming the off-resonance condition $\omega\not=\omega_0$).

In 1981 Witten \cite {WE81} introduced his non-relativistic model of supersymmetry.
The one-particle Hamiltonian reads
\be
H=-\frac {\hbar^2}{2m}\frac {{\rm d}^2\,}{{\rm d}x^2}+\frac 12mw^2(x)
+\frac 12\hbar \sigma_z\frac {{\rm d}w}{{\rm d}x}.
\label {SUSYham}
\ee
It can be written as
\be
H=\{Q,Q^\dagger\}
\label {superchargerel}
\ee
with
\be
Q=\frac 1{\sqrt{2m}}\left(P-im w(x)\right)\sigma_+,
\ee
where
\be
P=\frac {\hbar}i\frac {{\rm d}\,}{{\rm d}x}
\quad\mbox{ and }\quad
\sigma_\pm=\frac 12(\sigma_x\pm i\sigma_y).
\ee
It is straightforward to observe that this supersymmetric
Hamiltonian satisfies (\ref {model:lesstriv}) with $M=\sigma_x$ and $\gamma=2$.

A highly non-trivial example of (\ref {model:ccr}) is the
transverse Ising model \cite {PP70}, with Hamiltonian
\be
H&=&-\frac 12\sum_{ij}J_{ij}\sigma_i^z\sigma_j^z-h\sum_k\sigma_k^x.
\ee
Assume that $J_{ij}=J_{ji}$ and $J_{ii}=0$. Choose $M=\sum_k\sigma_k^x$.
Then one calculates
\be
[H,M]&=&-2i\sum_{ij}J_{ij}\sigma_i^y\sigma_j^z,\\
\strut
[[H,M],M]&=&8(H+hM)+4\sum_{ij}J_{ij}\sigma_i^y\sigma_j^y.
\ee
The relation between $[[H,M],M]$ and $H$ is not so easy to analyse as
in the previous examples. However, some further calculation shows that (\ref {model:ccr})
is satisfied with $\gamma=4$.

Finally consider the Hubbard model. The Hamiltonian is % \cite {HJ63}
\be
H=-\sum_{i,j}t_{ij}\sum_{\sigma=\uparrow,\downarrow}b^\dagger_{i,\sigma}b_{j,\sigma}
+\alpha\sum_kn_{k,\uparrow}n_{k,\downarrow}
\label {model:hubbard}
\ee
with $b^\dagger_{i,\uparrow}$ and $b^\dagger_{i,\downarrow}$ the creation operators
for an electron with spin up respectively spin down located at site $i$,
and with $n_{i,\sigma}=b^\dagger_{i,\sigma}b_{i,\sigma}$ the particle
counting operator at site $i$ and spin $\sigma$.
The coefficients $t_{ij}$ and $\alpha$ are assumed to be real.
They must satisfy $t_{ij}=t_{ji}$ to make $H$ Hermitean.
In addition, one assumes that $t_{ii}=0$. Now let
\be
M&=&\sum_kn_{k,\uparrow}n_{k,\downarrow}.
\ee
Then, calculations similar to those of the transverse Ising model
show that (\ref {model:ccr}) is satisfied with $\gamma=1$.
In this case $M$ counts the number of electron pairs
sharing a lattice site.

The above examples make it clear that it is worthwhile to investigate the relation (\ref {model:ccr}).
Its first consequence is that one can write the Hamiltonian $H$ into the form
\be
H=H_{\rm ref}+R+R^\dagger,
\label {model:hdecomp}
\ee
with $H_{\rm ref}$ and $R$ satisfying
\be
[H_{\rm ref},M]&=&0
\label {model:Mcond1}\\
\strut
[R,M]&=&\gamma R.
\label {model:Mcond2}
\ee
Indeed, let
\be
H_{\rm ref}&=&H-\frac 1{\gamma^2}[[H,M],M],\\
R&=&\frac 1{2\gamma^2}[[H,M],M]+\frac 1{2\gamma}[H,M].
\label {model:KRdef}
\ee
It is then straightforward to check that (\ref {model:hdecomp}, \ref {model:Mcond1}, \ref {model:Mcond2})
are verified. Conversely, if a Hamiltonian $H$ can be written as (\ref {model:hdecomp}),
with $H_{\rm ref}$ and $R$ satisfying (\ref {model:hdecomp}, \ref {model:Mcond1}, \ref {model:Mcond2})
for some $M$ and $\gamma$ then (\ref {model:ccr}) follows automatically.
Hence, (\ref  {model:ccr}) and (\ref {model:hdecomp}, \ref {model:Mcond1}, \ref {model:Mcond2})
are equivalent. The algebraic relation (\ref {model:Mcond2}) is also satisfied in Fock space by the
annihilation operators and the particle counting operator. This is the motivation to call
$M$ the counting operator.

If $M=E$ is an orthogonal projection operator then one has
\be
H_{\rm ref}&=&EHE+(\Io-E)H(\Io-E),\crcr
R&=&(\Io-E)HE.
\ee
This means that $H_{\rm ref}$ is the diagonal part in the block matrix representation
determined by $E$, while $R$ and $R^\dagger$ are the off-diagonal contributions.
Note that in this case $RR^\dagger+R^\dagger R=-[H,E]^2$. Hence, if $[H,E]$
is a multiple of one then $R$ and $R^\dagger$ satisfy anti-commutation relations.

A simple calculation shows that $H_{\rm ref}=0$ and $R=\frac 14\hbar\omega_0(\sigma_z+i\sigma_y)$
in the example of the Pauli spin,
$H_{\rm ref}=M+\zeta^2/\hbar\omega$ and $R=\zeta b-\zeta^2/\hbar\omega$ for the harmonic oscillator, and
$H_{\rm ref}=H_0$ and $R=\hbar\kappa b^\dagger\sigma_-$ in the Jaynes-Cummings model.
In the supersymmetric model is
\be
H_{\rm ref}&=&-\frac {\hbar^2}{2m}\frac {{\rm d}^2\,}{{\rm d}x^2}+\frac 12mw^2(x),\\
R&=&\frac 14\hbar \frac {{\rm d}w}{{\rm d}x}\,(\sigma_z+i\sigma_y).
\ee
In the transverse Ising model is
\be
H_{\rm ref}&=&-\frac 14\sum_{i,j}J_{i,j}\left(\sigma_i^y\sigma_j^y+\sigma_i^z\sigma_j^z\right)
-h\sum_k\sigma_k^x,\\
R&=&\frac 18\sum_{i,j}J_{i,j}\left(\sigma_i^y\sigma_j^y-\sigma_i^z\sigma_j^z\right)
+\frac i4\sum_{i,j}J_{i,j}\sigma_i^y\sigma_j^z.\crcr
& &
\ee
Note that $H_{\rm ref}$ is the Hamiltonian of the quantum XY-model.
In the Hubbard model is
\be
H_{\rm ref}&=&-\sum_{i,j}t_{ij}b^\dagger_{i,\uparrow}b_{j,\uparrow}(1-(n_{i\downarrow}-n_{j\downarrow})^2)\crcr
& &
-\sum_{i,j}t_{ij}b^\dagger_{i,\downarrow}b_{j,\downarrow}(1-(n_{i\uparrow}-n_{j\uparrow})^2)\crcr
& &+\alpha\sum_kn_{k,\uparrow}n_{k,\downarrow},\\
R&=&-\sum_{i,j}t_{ij}b^\dagger_{i,\uparrow}b_{j,\uparrow}(1-n_{i\downarrow})n_{j\downarrow}\crcr
& &
-\sum_{i,j}t_{ij}b^\dagger_{i,\downarrow}b_{j,\downarrow}(1-n_{i\uparrow})n_{j\uparrow}.
\ee

%%%%%%%%%%%%%%

Another consequence of the assumption (\ref {model:ccr}) is that the spectrum of $M$
necessarily consists of equally spaced energy levels. Indeed, from $M\psi=\mu\psi$ follows
\be
M(R\psi)=R(M-\gamma)\psi=(\mu-\gamma)(R\psi).
\ee
Hence, either $R\psi=0$ or $\mu-\gamma$ is also an eigenvalue of $M$.
Similarly, one concludes that either $R^\dagger\psi=0$ or $\mu+\gamma$ is also an eigenvalue of $M$.

Creation and annihilation operators may be used to generate a series of eigenstates starting
from a single eigenstate, which has been found by other means.
Let us show that the operators $R$ and $R^\dagger$ introduced above can serve the same purpose.

Introduce the notion of a {\sl stable eigenstate} \cite {foot2} as an eigenstate $\psi$ of $H$
for which either $R\psi=R^\dagger\psi=0$ or
for which numbers $x,y$ exist so that
\be
(xR+yR^\dagger)\psi=\psi.
\label {stable}
\ee
The exceptional case that $R\psi=R^\dagger\psi=0$ is added for convenience.
Both eigenstates of the Pauli spin Hamiltonian are stable, with $x=y=\pm 2/\hbar\omega_0$.
The ground state $|0\rangle$ of the harmonic oscillator is stable. Indeed,
one has $R|0\rangle=-(\zeta^2/\hbar\omega)|0\rangle$. Hence, (\ref {stable})
is satisfied with $x=-\hbar\omega/\zeta^2$ and $y=0$.

Given a stable eigenstate $\psi$ of $H$ with eigenvalue $\nu$
one can try to construct a new eigenstate $\chi$ defined by $\chi=f(M)\psi$.
One has
\be
H\chi
&=&[H,f(M)]\psi+\nu\chi\crcr
&=&[R+R^\dagger,f(M)]\psi+\nu\chi\crcr
&=&[f(M+\gamma)-f(M)]R\psi\crcr
& &+[f(M-\gamma)-f(M)]R^\dagger\psi+\nu\chi.
\ee
Now let $f(u)$ be a function such that for some real $\lambda$
\be
& &f(M+\gamma)R\psi+f(M-\gamma)R^\dagger\psi\crcr
& &=(1+\lambda x)f(M)R\psi\crcr
& &\quad +(1+\lambda y)f(M)R^\dagger\psi.
\label {eigen:fdef}
\ee
Then the expression becomes, using (\ref {stable}),
\be
H\chi
&=&\lambda f(M)\left(xR+yR^\dagger\right)\psi+\nu\chi\crcr
&=&(\nu+\lambda)\chi.
\ee
One concludes that $\chi$ is an eigenstate of $H$ with eigenvalue $\nu+\lambda$.
The construction of new eigenstates of $H$ is therefore reduced to finding functions $f(u)$
solving the eigenvalue equations (\ref {eigen:fdef}).

As a first application let us assume that $\psi$ is a stable eigenstate with $xy\not=0$
and $x+y\not=0$.
Then there exists at least one other stable eigenstate $\chi$ of the form $\chi=f(M)\psi$.
Indeed, make the choice $f(u)=\exp(zu)$ with $z$ a complex constant. Then
a sufficient condition for (\ref {eigen:fdef}) to hold is
\be
e^{zM}e^{z\gamma}R\psi&=&(1+\lambda x)e^{zM}R\psi,\crcr
e^{zM}e^{-z\gamma}R^\dagger\psi&=&(1+\lambda y)e^{zM}R^\dagger\psi.
\ee
These equations have a trivial solution when
\be
\lambda=-\frac {x+y}{xy}.
\ee
The solution is obtained for $z$ satisfying $\exp(z\gamma)=-x/y$.
The resulting eigenstate $\chi$ is orthogonal to $\psi$ because the assumption $x+y\not=0$
implies $\lambda\not=0$ so that $\psi$ and $\chi$ have different eigenvalues.
It is straightforward to verify that $\chi$ is again a stable eigenstate.
However, repeating the above argument starting with $\chi$ reproduces the eigenstate $\psi$.
Hence, with this choice of $f(u)$ only one additional eigenstate can be obtained.

When applied to the Pauli spin example the above reasoning allows to derive one
of the two eigenstates from the other. The value of $\lambda$ is then $\pm \hbar\omega_0$.
Also all eigenvectors of the Jaynes-Cummings model are stable. It is well-known that
the spectrum of this model consists of a singlet, which is the ground state level,
and an infinity of doublets. The ground state is stable in a trivial way because
it satisfies $R\psi=R^\dagger\psi=0$. Each of the doublets consists of two stable
eigenvectors which transform into each other by the above mechanism.
In the one-dimensional Hubbard model with 4 sites and a half filled band
one can show \cite {AVN08} that several
pairs of stable eigenstates occur which transform into each other by the above mechanism.
In addition, these states are eigenstates of the anti-commutator $\{R,R^\dagger\}$.
Similarly, in the one-dimensional transverse Ising model with 3 and that with 4 sites
one can show that all eigenstates are stable and that they either
satisfy $R\psi=R^\dagger\psi=0$ or they occur in pairs which transform into
each other in the way described above \cite {NJ08u}.

The ground state of the harmonic oscillator does not satisfy the condition $xy\not=0$.
Hence, the previous result cannot be used to construct a second stable eigenstate,
in agreement with the observation that the ground state is the only stable eigenvector of the model.
In this case, no new eigenstates of the form $f(M)|0\rangle$ can be obtained from the
ground state $|0\rangle$, as can be seen immediately using the commutation
relation $[M,b]=-\hbar\omega b$. However, it is well known that all eigenstates
are obtained by repeated action of the creation operator $b^\dagger=R^\dagger/\zeta+\zeta/\hbar\omega$.

More generally, if in some model an eigenvector $\psi$ of $H$ exists such that $xR\psi=\psi$
then an annihilation operator $B$, annihilating $\psi$,
can be defined by $B=R-x^{-1}$. However, in general not much is known
about the commutators $[B,B^\dagger]=[R,R^\dagger]$ and $[H_{\rm ref},B^\dagger]$.
For this reason, no further progress was made in this case.

Once a pair of stable eigenvectors has been obtained one can try
to find functions $f$, other than exponential ones, satisfying (\ref {eigen:fdef}).
This does indeed work for the following artificial example.
Fix $\kappa$, $\xi$, $\mu$, and $\nu$ so that 
$\kappa\not=0$, $\mu\not=\xi$, and $\xi+\nu=2\mu$.
Let
\be
H=\left(\begin {array}{lcr}
         \xi &\kappa&0\\\kappa&\mu&\kappa\\0&\kappa &\nu
        \end {array}\right)
\ee
and
\be
M=\left(\begin {array}{lcr}
         1 &0&0\\0&0&0\\0&0 &-1
        \end {array}\right).
\ee
The eigenvectors and eigenvalues can be calculated explicitly.
Two of the eigenvectors are stable. They are of the form
$\psi=(a^2,2a,2)^{\rm T}$. The third eigenvector is of the form
$\chi=(1,\delta,-1)^{\rm T}$ and satisfies $\chi=f(M)\psi$,
with $f(u)$ of the form $f(u)=A+Bu$. One concludes that the
three eigenvectors together form a triplet. Note that this
example of 3-by-3 matrices has been considered in \cite {YAS02} as well.
One can write $H=H_{\rm ref}+\kappa V$ and $I=M+\kappa W$, with $\kappa V=R+R^\dagger$.
However, there does not exist a $\kappa$-independent matrix $W$ such that $H$ and $I$
commute for all values of $\kappa$. Hence, in the terminology of \cite {YAS02},
the pair $(H,I)$ is not integrable.
Similar triplets exist in the 4-site Hubbard ring
--- see \cite {AVN08}.

{\sl In summary,} we have studied model Hamiltonians $H$ together with a Hermitean operator $M$
such that the third commutator of $H$ with $M$ is proportional to the first commutator.
Then the spectrum of $M$ consists of equidistant levels and $M$ is called the counting operator.
In this context one can define operators $R$ and $R^\dagger$ which have some similarity
with annihilation and creation operators.  An eigenvector $\psi$ of $H$ is said to be stable
if either $R\psi=R^\dagger\psi=0$ or there exist $x,y$ such that $(xR+yR^\dagger)\psi=\psi$.
If $xy\not=0$ and $x+y\not=0$ then a new stable eigenvector can be constructed whose
eigenvalue is decreased with $(x+y)/xy$. This is the main result of the present paper.
But the analysis also suggests to decompose the spectrum of $H$ into multiplets
determined by the symmetries of the reference Hamiltonian $H_{\rm ref}=H-R-R^\dagger$.
This view is supported by the detailed analysis of some small
systems, including the 4-site Hubbard ring, which is studied in \cite {AVN08}.

{\bf Acknowledgements} We are grateful to Prof.~J.~Perk
for pointing out \cite {DG82,PJ89,DB90}, and to Prof.~M.S.~Plyushchay
for pointing out \cite {KP02}.

%%%%%%%%%%%%%%%%%%%%%%%%%%%%%%%%%%%%%%%%%%%%%%%%%%%%%%%%%%%%%%%%%%%%%%%%%%%%%%%%%


\newpage
\begin{thebibliography}{99}

\raggedright\parskip 0pt

\bibitem {KD65}
D.H. Kobe,
% {\sl Criteria for the best Bogoliubov quasiparticle,}
Phys. Rev. {\bf 140}, A825 -- A829 (1965).

\bibitem {WC08}
B.W. Williams and T.C.J. Celius,
% {\sl New Potentials for Old: The Darboux Transformation in Quantum Mechanics,}
J. Chem. Educ. {\bf 85}, 576 -- 584 (2008).

\bibitem {KB66}
B. Kaufman, J. Math. Phys. {\bf 66}, 447 -- 457 (1966).

\bibitem {FLP01} W. Garcıa Fuertes, M. Lorente and A.M. Perelomov,
% {\sl An elementary construction of lowering and raising
%operators for the trigonometric Calogero–Sutherland model,}
J. Phys. A{\bf 34}, 10963 -- 10973 (2001).

\bibitem {SR02} R. Schumann,
% {\sl Thermodynamics of a 4-site Hubbard model by analytical diagonalization,}
Ann. Phys. (Leipzig) {\bf 11}, 49 -- 87 (2002).

\bibitem {AVN08}
T. Verhulst, B. Anthonis, and J. Naudts, arXiv:0811.3077.

\bibitem {LW68}
E. H. Lieb and F. Y. Wu, Phys. Rev. Lett. {\bf 20}, 1445 -- 1448
(1968).

\bibitem {LW03}
E.H. Lieb, F.Y. Wu,
Physica A {\bf 321}, 1 -- 27 (2003).

\bibitem {HL71}
O.J. Heilmann and E.H. Lieb,
Ann. N.Y. Acad. Sci. {\bf 172}, 584 -- 617 (1971).

\bibitem {NW29}
J. von Neumann and E. Wigner,
%{\sl Ueber das Verhalten von Eigenwerte bei adiabatische Prozessen,}
Physik. Zeitschr. XXX, 467 -- 470 (1929).

\bibitem {YAS02}
E.A. Yuzbashyan, B.L. Altshuler and B.S. Shastry,
J. Phys. A {\bf 35}, 7525 -- 7547 (2002).

\bibitem {foot1}
This relation, when applied to the two parts of a self-dual Hamiltonian,
is known as the Dolan-Grady condition \cite {DG82}.
It is related with the Onsager algebra and with super\-integrability \cite {PJ89,DB90}
and has been used in the context of supersymmetry \cite {KP02}.

\bibitem {RJC99}
A.K. Rajagopal, K.L. Jensen, F.W. Cummings,
% {\sl Quantum entangled supercorrelated states in the Jaynes-Cummings model,} 
Phys. Lett. A{\bf 259}, 285 -- 290 (1999)

\bibitem {BHN02}
Y. B\'erub\'e-Lauzi\`ere, V. Hussin, and L. M. Nieto,
% {\sl Annihilation operators and coherent states for the Jaynes-Cummings model,}
Phys. Rev. A {\bf 305}, 135 -- 143 (2002).

\bibitem {WE81}
E. Witten,
% {\sl Dynamical breaking of supersymmetry,}
Nucl. Phys.  B {\bf 188}, 513 -- 554 (1981).

\bibitem {PP70}
P. Pfeuty,
% {\sl The one-dimensional Ising model with a transverse field,}
Ann. Phys. {\bf 57},  79 -- 90 (1970).

\bibitem {foot2}
Alternatively, stable states could be called weakly coherent,
motivated by the property of coherent states being
eigenvectors of the annihilation operator.
However, the presence of a genuine coherence property is not immediately clear.

\bibitem {NJ08u} J. Naudts, unpublished.

\bibitem {DG82}
L. Dolan and M. Grady,
% {\sl Conserved charges from self-duality,}
Phys. Rev. D {\bf 25}, 1587 -- 1604 (1982).

\bibitem {PJ89} J.H.H. Perk,
% {\sl Star-triangle equations, quantum Lax pairs, and higher genus curves,}
in: {\sl Proc. 1987 Summer Research Institute on Theta Functions,}
Proc. Symp. Pure Math. {\bf 49}, part 1, (Am. Math. Soc., Providence, R.I., 1989), 341 -- 354.

\bibitem {DB90} B. Davies,
% {\sl Onsager's algebra and superintegrability,}
J. Phys. A {\bf 23}, 2245 -- 2261 (1990).

\bibitem {KP02}
S.M. Klishevich and M.S. Plyushchay,
Nucl.Phys.B {\bf 628}, 217 -- 233 (2002).


\end {thebibliography}
\end{document}